\documentclass[9pt,twocolumn,twoside]{opticajnl}
\journal{opticajournal} 
\setboolean{shortarticle}{false}
\usepackage{lineno}
\usepackage{braket}
\usepackage{amsmath}

\usepackage{multirow}
\usepackage{float}

\title{High-dimensional quantum key distribution with resource-efficient detection}

\author[1,6,*]{Maciej Ogrodnik}
\author[1,6]{Adam Widomski}
\author[2]{Dagmar Bru\ss{}}
\author[3]{Giovanni Chesi}
\author[4]{Federico Grasselli}
\author[2]{Hermann Kampermann}
\author[3]{Chiara Macchiavello}
\author[5]{Nathan Walk}
\author[2]{Nikolai Wyderka}
\author[1]{Micha{\l} Karpi\'nski}
\affil[1]{Faculty of Physics, University of Warsaw, Pasteura 5, 02-093 Warszawa, Poland}
\affil[2]{Institut f\"ur Theoretische Physik III, Heinrich-Heine-Universit\"at D\"usseldorf, Universit\"atsstra\ss{}e 1, D-40225 D\"usseldorf, Germany\\}
\affil[3]{National Institute for Nuclear Physics, Sezione di Pavia, Via Agostino Bassi 6, 27100 Pavia, Italy}
\affil[4]{Leonardo Innovation Labs -- Quantum Technologies, Via Tiburtina km 12,400, 00131 Rome, Italy}
\affil[5]{Dahlem Center for Complex Quantum Systems, Freie Universit\"at Berlin, Arnimallee 14, D-14195 Berlin, Germany}

\affil[6]{These authors contributed equally to this work.}
\affil[*]{mm.ogrodnik@uw.edu.pl}

\begin{abstract}
While quantum key distribution (QKD) based on two-dimensional (qubit) encoding is a mature, field-tested technology, its performance is lacking for many cryptographic applications.
High-dimensional encoding for QKD enables increased achievable key rates and robustness as compared to the standard qubit-based systems.
However, experimental implementations of such systems are more complicated, expensive, and require careful security analysis as they are less common.
In this work we present a proof of principle high-dimensional time-phase BB84 QKD experiment using only one single-photon detector per measurement basis.
We employ the temporal Talbot effect to detect QKD symbols in the control basis, and show experimentally-obtained simplistic key rates for the two-dimensional and four-dimensional case, including in an urban fiber network.
We present a comparison of a simplistic secret key rate obtained from a standard security proof with the one derived from a recently devised proof using a tunable beam splitter to display security issues stemming from asymmetric detection efficiencies in the two bases.
Our results contribute to the discussion of the benefits of high-dimensional encoding and highlight the impact of security analysis on the achievable QKD performance. 
\end{abstract}

\setboolean{displaycopyright}{false}

\begin{document}

\maketitle

\section{Introduction}
Advancements in quantum computing pose a significant threat to the security of communication systems.
Quantum key distribution (QKD) provides protocols with provable security under minimal assumptions on the adversary, in contrast to classical cryptography solutions \cite{Aaronson2016, qcrypto-in-algorithmica}.
The security of practical QKD systems relies on how well the system corresponds to the theoretical model of a QKD protocol \cite{Jain-QKD-attacks, Sajeed2021}.
Different physical realizations of QKD protocols come from practical considerations such as cost, speed or complexity of the system \cite{qkd2020review, advancesQKD2020}.
Many systems use qubits (states in a 2-dimensional Hilbert space) which carry one bit of information per photon \cite{Grunenfelder2023, Li2023, Yuan18, secoqc}.
With $d$-dimensional quantum states (qudits), one can increase the bits of information per symbol to $\log_2 (d)$ bits \cite{vagniluca2020efficient, Yu2025}.
Qudits enable tolerating higher quantum bit error rates (QBERs) present in quantum communication \cite{Cerf2002, sheridan2010} and overcoming problems related to saturation of the receiver's single-photon detectors \cite{Islam2017}.
As a result, higher secret key rate values can be achieved in comparison to the binary encoding ($d=2$) scenario \cite{vagniluca2020efficient, hdqkd2019review, Sit:17}.
Constructing a high-dimensional QKD system requires the ability to generate and detect high-dimensional quantum states and their superpositions
in the key generation and control bases, which comes at a cost of higher complexity as compared to qubit systems. 
Various photons' degrees of freedom were used in high-dimensional QKD experiments.
Experiments using OAM \cite{Bouchard2018experimental, mirhosseini2015high, cozzolino2019orbital}, polarization-OAM hybrid \cite{wang2019characterizing}, or spectral-temporal degree of freedom were successfully performed over fiber \cite{Zhong2015, Kues2017, Lee19} and free-space \cite{steinlechner2017distribution} quantum channels.

Most security proofs for QKD are based on abstract quantum states, independent of their physical realization. To apply these proofs to real-world systems, it is necessary to ensure that the setup reliably prepares and measures quantum states according to the assumptions of the proof, even in the presence of adversarial manipulations.
The time-phase variant of the BB84 protocol with weak coherent pulses and the decoy state method is a well-researched encoding scheme \cite{boaron2018secure, Islam2017, Jin:19, qkd2020review}.
Typically, Franson interferometers \cite{franson1989bell, Zhong2015} are employed to detect phases between light pulses.
However, a single interferometer measures the phase between only two time bins.
To detect high-dimensional states such interferometers have to be nested \cite{islam2017robust, brougham2013security, Yu2025}, which implies higher complexity, cost, and losses.
The probability to detect a qudit with a passive interferometer tree is scaled by the $1/d$ factor due to the increased number of paths the superposition may take \cite{islam2017robust, widomski2024}. The alternative approach of active switching enables more efficient detection of time-bin superposition, at the cost of significantly increased experimental complexity \cite{Vedovato2018, Wang2018}.

Alternatively, temporal mode superpositions were measured by means of the optically nonlinear quantum pulse gate \cite{Eckstein2011, Reddy2018, serino2023}.
However, these methods necessitate active spectral modification, involving complex setups and ultrashort laser pulses at the detection stage.
In principle, tailored time-frequency basis transformations could be used \cite{Ashby20, Joshi22, Karpinski2021}, but their efficient experimental implementation is still to be developed.
Additionally, such implementations often result in a mismatch between measurement probability in the two bases.
This inequality should be considered in the security proof and relevant key rate calculations.
Several attempts to incorporate this effect have been made \cite{trushechkinmismatch1, trushechkinmismatch2, blockdiagonalassumption, detectionefficiencymismatchLo, detectionefficiencymismatchLydersen, detectionefficiencymismatchMa}, yet it is a major challenge to fully model the real operation of physical devices such that all side-channel attacks can be avoided.

In this work we present an alternative scheme for high-dimensional QKD using time-phase encoding and spectral-temporal decoding, which allows using the same passive receiver architecture regardless of the dimension.
We use the temporal Talbot effect \cite{talbot1836, jannson1981temporal, widomski2024}, to discriminate qudits in the control basis.
Our experimentally simple method only requires a dispersive medium and a single time-correlated single photon counter.
Qudit exchange was tested in two and four-dimensional scenario for fiber-based quantum channel attenuations spanning up to $27.24$~dB using the same setup.
The measurements were conducted utilizing in-lab fibers as well as urban dark-fiber infrastructure of the University of Warsaw.

As noted in \cite{theorypaper} spectral-temporal decoding may result in mode-dependent detection efficiency in violation of the assumptions of standard protocol security proof.
That is why in addition to the novel protocol implementation and its tests, we present a theoretical comparison of the asymptotic secret key rates for different dimensions considering experimentally-obtained parameters using two different security proofs.
In particular, we compare the \emph{simplistic} key rates obtained with the standard BB84 protocol, which assumes that the detection probability at Bob is independent of the measurement basis, and with a recent security proof by  F.~Grasselli {\em{et.~al.}}~\cite{theorypaper}, where the asymmetric detection efficiency in the two bases is accounted for.

The manuscript is organized as follows: in the ``Protocol'' section we describe our encoding scheme, detection method and show the formula for the simplistic key rate.
In the``Experiment'' section we describe the setup, including the locations of fiber network nodes, and data acquisition methods.
The results, QBER, and theoretical simplistic key rate analysis using the standard high-dimensional BB84 approach are contained in the ``Results and discussion'' section.
We proceed with a short description of vulnerabilities in the ``Security discussion'' section. We then apply the new security proof in the ``Tunable beam splitter protocol'' section to highlight the differences stemming from security analysis, and finally conclude and compare the results in the ``Conclusion'' section.

\section{Protocol}

Let us consider optical pulses in $d$ orthogonal time bins $\ket{t_0}$, $\ket{t_1}$,..., $\ket{t_{d-1}}$ forming the $Z$ (key generation) basis, and $d$ superpositions of those states that form a discrete Fourier transform of the $Z$ basis \cite{barnett2009quantum}, comprising the $X$ (control) basis:

\begin{equation}
    \ket{f_n} = \frac{1}{\sqrt{d}} \sum_{m=0}^{d-1} e^{\frac{-2\pi i n m}{d}} \ket{t_m}.
    \label{eq:dft-basis}
\end{equation}
Then, the two bases are mutually unbiased and satisfy the relation \cite{grasselliqkd}:
\begin{equation}
    |{\braket{f_n|t_m}}|^2 =\frac{1}{d} \quad\forall\, m,n.
    \label{eq:mub}
\end{equation}
We use those bases to realize the $d$-dimensional BB84 protocol with information encoded in the time-phase degrees of freedom. We utilize the two-decoy scheme to avoid photon-number splitting attacks.

\begin{figure}[h!]
\centering
\includegraphics[scale=1.0, width=8.5cm]{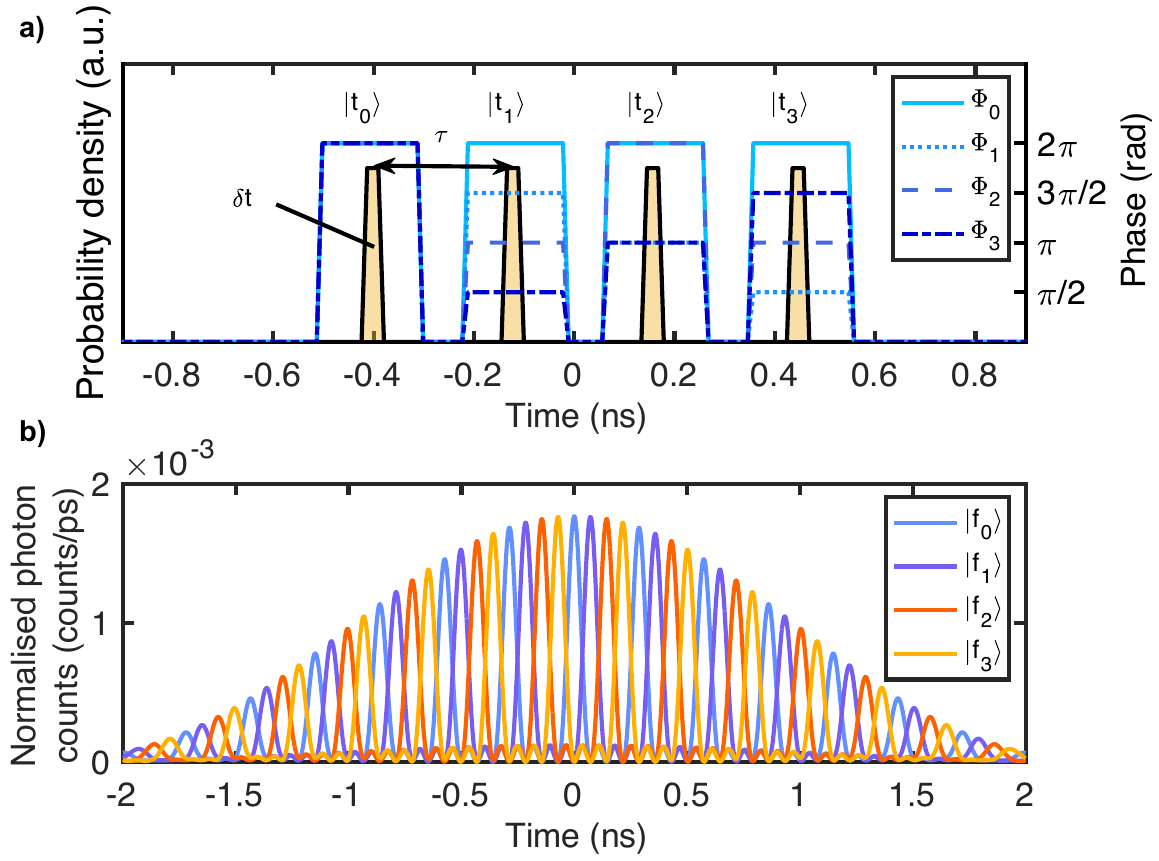}
\caption{State preparation.
a) Generating high-dimensional states.
The symbols were approximately rectangular optical pulses attenuated to single-photon level, 46 ps wide, and separated by 284 ps.
Electrical signals used to impose phase modulation were 150 ps wide, and their amplitude was adjusted with respect to the half-wave voltage of the employed phase modulator, cf. Fig.\ \ref{fig:setup}.
b) Simulated temporal Talbot effect on time of arrival probability density functions of high-dimensional states generated as in a).
\label{fig:generation}}
\end{figure} 

Fig.~\ref{fig:generation}a) depicts the temporal profiles of optical and electrical signals used for generating the quantum states used in this work for the $d=2$ and $d=4$ case.
We adjust the separation between the components of the superposition $\tau$ and their width $\delta t$ such that detection using the temporal Talbot effect is possible.
The measurement is realized by means of temporal self-imaging using a dispersive medium with a specific group delay dispersion (GDD), further called the Talbot GDD \cite{jannson1981temporal}.
The temporal location of the self-images depends on the relative phases of superposition's components, which makes different superpositions distinguishable \cite{widomski2024}. Therefore, information in both the bases is measured only in the time domain.
The temporal Talbot condition imposes the following relation between the pulse separation ($\tau$) and GDD  ($\beta_2 $):
\begin{equation}
    \tau=\sqrt{\frac{2\pi\beta_2}{s}},\, s = 1,2\ldots\ .
     \label{eq:TalbotGDD}
\end{equation}
This pulse separation affects the detection error rate as discussed in \cite{widomski2024}.
An attacker might influence the pulse separation and increase the measured QBER.
Yet, it does not change the probability of detection, hence it does not open the detection asymmetry loophole discussed in \cite{theorypaper}.

By employing the temporal Talbot effect we are able to construct an all-fiber detection system with one single-photon detector per measurement basis, resulting in a very simple setup.
We show how the detection error related to our method scales with the dimension in the "Results and Discussion" section. 
In section \ref{sec:discussion} we discuss potential security issues connected with that detection method. It could be argued that the effect of those issues could be somewhat mitigated by spectral filters. Here we take a different approach. We analyze the security according to the standard $d$-dimensional BB84 proof to get simplistic key rates and later in the  section \ref{sec:tbs} we discuss modifications that eliminate the relevant class of detection vulnerabilities.

The asymptotic secret key rate for a $d$-dimensional BB84 protocol in the 2-decoy variant is given by \cite{theorypaper}:  
\begin{align}
   & r_{\rm BB84} = p_Z^2 {\textstyle\sum_{j=1}^3} p_{\mu_j} \left\lbrace e^{-\mu_j} \underline{Y^Z_{0}} \log_2 d  \right.\nonumber\\
        &\quad\left. + e^{-\mu_j} \mu_j \underline{Y^Z_{1}} \left[\log_2 d - u(\,\overline{e_{X,1}}\,)\right] - G^Z_{\mu_j} u(Q_{Z,\mu_j}) \right\rbrace,
        \label{eq:skr}
\end{align}
where $p_Z$ is the probability to prepare (measure) in the $Z$ basis, while the $X$ basis is chosen with probability $1-p_Z$.
Alice selects the symbol $j\in\{0,\dots,d-1\}$ ($k\in\{0,\dots,d-1\}$)  and intensity $\mu_i \in\mathcal{S}:=\{\mu_1, \mu_2, \mu_3\}$ with probabilities $p_{\mu_1}$, $p_{\mu_2}$ and $p_{\mu_3}=1-p_{\mu_1}-p_{\mu_2}$.
An upper bound on the $X$ basis QBER is denoted as $\overline{e_{X,1}}$, $\underline{Y^Z_{0}}$ and $\underline{Y^Z_{1}}$ are the lower bounds on 0 and 1-photon yields in the Z-basis, $ G^Z_{\mu_j}$ and $Q_{Z,\mu_j}$ are respectively the gain and QBER corresponding to a state with $\mu_j$ mean photon number.

The function $u(x)$ appearing in Eq.~\eqref{eq:skr} is defined as:
\begin{align}
    u(x)= \left\lbrace \begin{array}{ll}
       h(x) + x \log_2 (d-1)  & \mbox{if } x\in\left( 0,1-\frac{1}{d}\right)  \\[1ex]
        \log_2 d &  x \in \left[1-\frac{1}{d},1\right),
    \end{array}\right.
    \label{u(x)}
\end{align}
with $h(x)=-x \log_2 x -(1-x)\log_2 (1-x)$ the binary entropy. 
For the asymptotic key rate the number of test rounds is negligible, so we have $p_Z = 1$ and $p_{\mu_1} = 1$. 
To determine key rate values we experimentally measure gains and QBERs for every $\mu$ value in the $X$ and $Z$ bases.
Those values are further used to calculate the bounds on yields and phase error rate (cf. Appendix E 1 in \cite{theorypaper}), based on which we finally compute the key rate value.
A detailed description of Eq.~\eqref{eq:skr} is provided in Section V A of ref.\ \cite{theorypaper}.

\section{Experiment} \label{setup}

\begin{figure}[h!]
\centering
\includegraphics[scale=1.0, width=8.5cm]{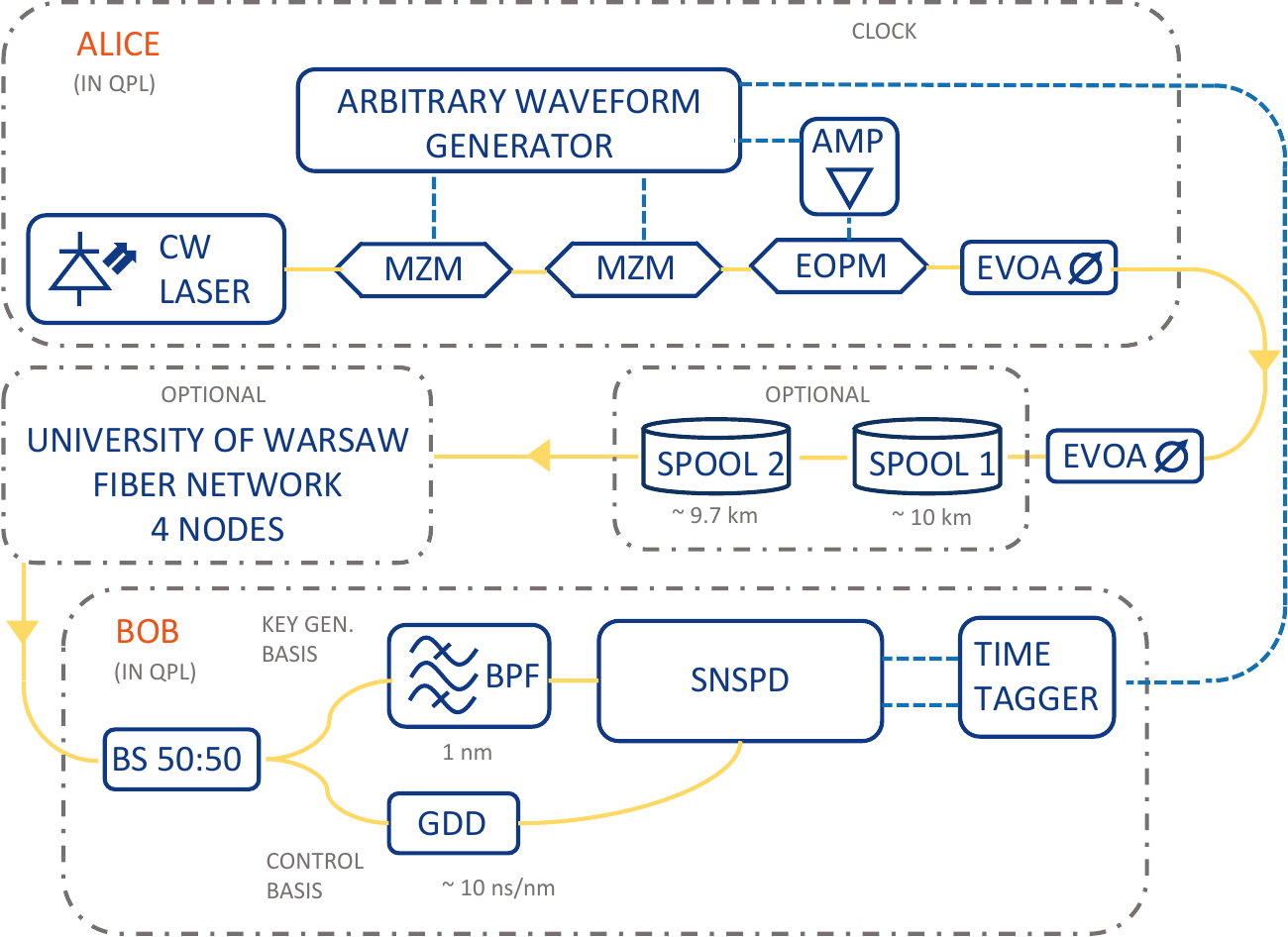}
\caption{Schematic of the key components of the experimental setup. A continuous wave (CW) laser is modulated with a Mach-Zehnder modulator (MZM) and phase modulator (EOPM) to generate optical pulses forming superpositions used as signals and decoy states. The optical signals are attenuated to a single photon level with a variable attenuator (EVOA).
Attenuation of the quantum channel is controlled with another EVOA.
Two spools of optical fiber (spool 1 and spool 2) were added to simulate propagation over a real link. Alternatively, the quantum channel comprised dark fibers deployed in the University of Warsaw urban fiber network, allowing measurements over certain fixed distances (see Fig.~\ref{fig:nodes}) by making a fiber loop.
The signals impinge on a switch which we use to further detect either directly in the key generation basis ($Z$ basis) or control basis ($X$ basis). In the $Z$ basis the signals are detected by time correlated single photon counting. In the $X$ basis, by means of the temporal Talbot effect performed with the dispersion compensating module (DCM) providing group delay dispersion (GDD).
Both Alice and Bob were placed in the Quantum Photonics Laboratory (QPL) at the Faculty of Physics, University of Warsaw. Yellow lines represent optical fiber connections and dashed blue lines depict RF cables. \label{fig:setup}}
\end{figure}

The schematic of the experimental setup is presented in Fig.\ \ref{fig:setup}. 

\begin{figure}[h!]
\centering
\includegraphics[scale=1.0, width=8.5cm]{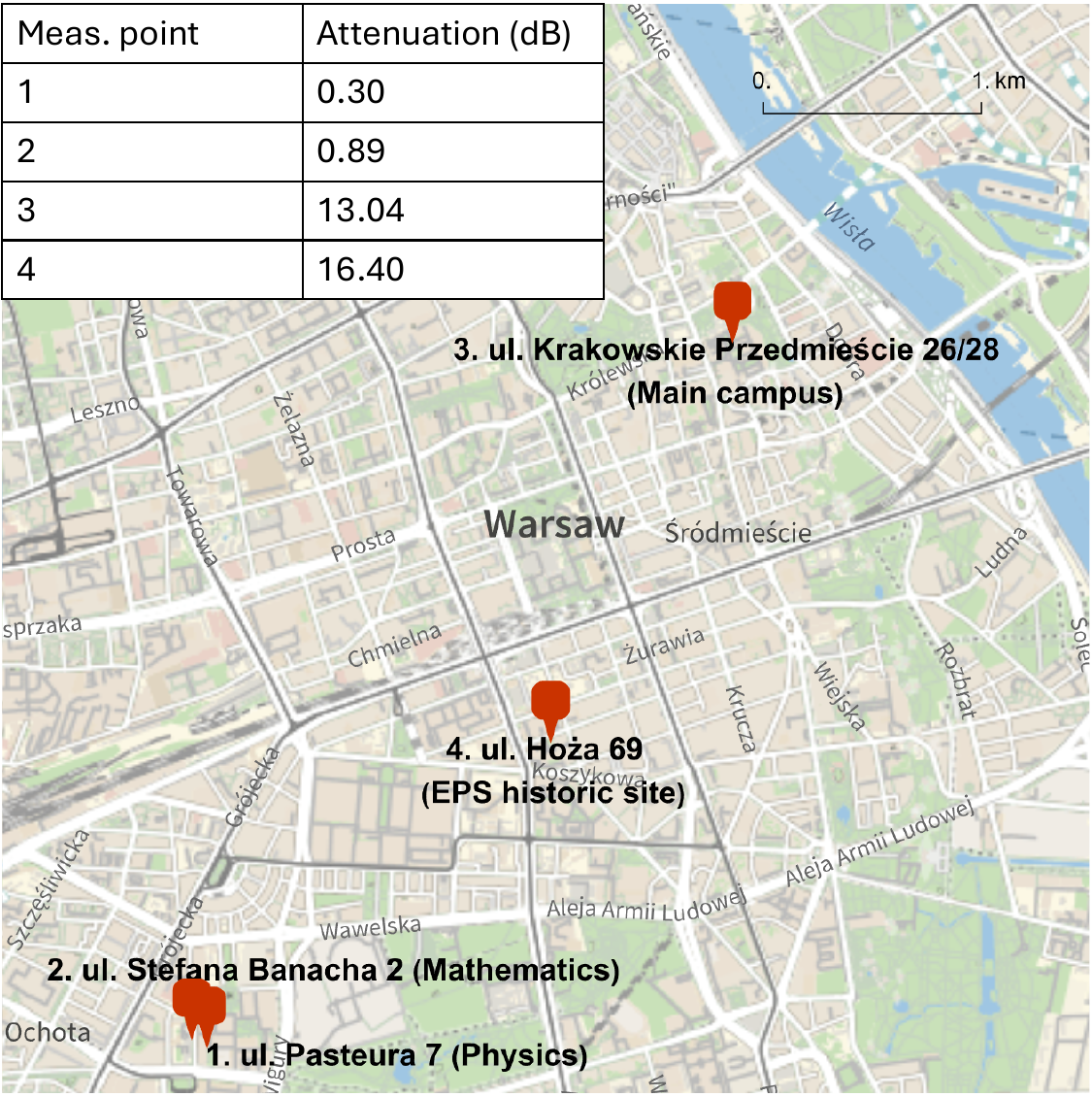}
\caption{Locations of the nodes of the dark fiber network of the University of Warsaw, overlaid on an Open Street Map 
of central Warsaw. Due to the network's architecture the signals were routed over a 13-km-long fiber link via the University's southern campus from node 2 to 3. Node 4 is located at a Historic Site of the European Physical Society (EPS).
\label{fig:nodes}}
\end{figure}

A distributed-feedback (DFB) laser operating at the telecommunication wavelength of $1560$~nm in the continuous wave (CW) mode was used as the main optical source. We amplified the optical signal with an erbium-doped fiber amplifier (EDFA, Pritel, HPP-PMFA-22-10) and employed a bandpass spectral filter to reduce the noise and provide sufficient power for a feedback loop, discussed later. 
The optical signals and decoy states for both the bases were generated by means of cascaded fast amplitude electro-optic modulation with a pair of Mach-Zehnder modulators (MZM, Thorlabs, LNA6213).
We biased both the MZMs for extinction with the direct current (DC) voltage from a stable power supply (Keysight, E36313A) using the feedback loop consisting of a 90:10 fiber optic beam splitter and a power meter (Thorlabs, PM400) (not shown in Fig.~\ref{fig:setup}). The second modulator was biased with another feedback loop comprising a separate CW laser for calibration, a power meter, an isolator, and a switch used to cut off the calibration laser for the time of key distribution. 
We controlled the mean photon numbers by driving both the modulators ($\mu_1$), one modulator ($\mu_2$), or none of them ($\mu_3$).
The mean photon numbers for in-laboratory experiments for dimension two and four were equal to $\mu_1 = 0.064 $, $\mu_2 = 0.684\cdot10^{-3} $, $\mu_3 = 177\cdot10^{-6}$ and $\mu_1 = 0.0617 $, $\mu_2 = 0.634\cdot10^{-3} $, $\mu_3 = 203\cdot10^{-6}$ respectively.
The mean photon numbers for infrastructural experiments for dimension two and four were equal to $\mu_1 = 0.236 $, $\mu_2 = 2.6\cdot10^{-3} $, $\mu_3 = 355\cdot10^{-6}$ and $\mu_1 = 0.2 $, $\mu_2 = 2.49\cdot10^{-3} $, $\mu_3 = 329\cdot10^{-6}$ respectively.

Subsequently, adequate phases (cf.\ Fig.~\ref{fig:generation}) were applied to every component of the superposition with an electro-optic phase modulator (EOPM, EOspace).
We would like to point out that the EOPM can also be used for phase randomization to eliminate coherence between subsequent rounds \cite{wang2005beating, rusca2018finite}.
All of the modulators support $40$~GHz of usable analog bandwidth. The radio-frequency (RF) driving  signals were  generated with a fast arbitrary waveform generator (AWG, Keysight, M1896A) providing a sampling rate of up to $92.16$~GSa/s and $35$~GHz of analog bandwidth. The full-width at half maximum (FWHM) duration of a one-bit signal was therefore $\sim\! 12\textrm{~ps}$. The phase factors were adjusted using the EOPM by programming four driving voltage signals consisting of approximately $150$-ps-wide rectangular pulses (Fig.~\ref{fig:generation}a) such, that their amplitudes corresponded to fractions or multiples of the phase modulator half-wave voltage (${V_\pi}$).
$3$~samples of AWG memory were used to generate a single optical pulse.
The RF signals driving the EOPM were also amplified with a high-bandwidth amplifier (RFLambda RFLUPA01G31GHz). The AWG also served as a source of the $10$~MHz clock (CLK) signal distributed over an electrical cable to Bob.
Finally, weak coherent states were generated by attenuating optical symbols to the single-photon level with an electronic variable optical attenuator (EVOA, Thorlabs, EVOA1550F).

To demonstrate the feasibility of our approach, we used a pre-programmed random sequence of symbols occupying $510720$ samples of our AWG's memory.
We then transmitted the sequence for $1$ minute at every $\mu$ level, for different quantum channel attenuations.
For the in-laboratory measurements the attenuation was set with another EVOA up to $20$~dB.
For convenience, the measurements were performed with a beam splitter that distributes signal to two detectors, effectively adding $3$~dB of attenuation.
On top of that we added two approximately 10-km-long spools of single-mode fiber to simulate a real link, which added $4.24$~dB attenuation (including the insertion loss of the EVOA).
To further demonstrate the feasibility we performed measurements over deployed dark fiber infrastructure (cf.\ Fig.~\ref{fig:nodes}) by making a fiber loop reaching the desired destination and reflecting the signal back to the Quantum Photonics Laboratory.
We acquired data for every available node with and without the two fiber spools. 
We manually switched the measurement basis on the receiver side.
Niobium-nitride superconducting nanowire single-photon detectors were used for time-domain measurements (SNSPDs, Single Quantum).
Detector efficiency was equal to $84\%$ according to the calibration data provided by manufacturer.
Detection events were registered with a time-to-digital converter  set to $1$-ps-wide  bins (Swabian Instruments, Time Tagger Ultra).
The SNSPDs exhibited the jitter of $5\textrm{~ps}$ root mean square RMS and the time tagger of $10 \textrm{ ps}$.
Thus, the Bob's resultant jitter was $11\textrm{~ps}$ RMS.

To measure in the control basis, the photons were additionally transmitted through a chirped-fiber-Bragg-grating-based dispersion compensating module (DCM), resulting in dispersive temporal broadening of the pulses.
The measured insertion loss of the DCM was equal to $2.67$~dB, and constituted the main contribution to the detection loss in the $X$ basis.
The contribution of other sources of optical loss was below $0.2$~dB, primarily due to fiber connector losses. The DCM provided group delay dispersion (GDD) of $12900 \textrm{ ps}^2$ equivalent to $562$~km of SMF-28 fiber.
This amount of dispersion enabled observing the temporal Talbot effect \cite{widomski2024}, which we used to distinguish superpositions of the optical signals with different phases. The DCM also acted like a bandpass filter transmitting wavelengths in range $1558$--$1562$~nm. To suppress the noise originating from the fiber network, we added an additional bandpass spectral filter (Haphit FPBP, $1$~nm BW) in the $Z$ basis. 

\section{Results} \label{results}

\begin{figure}[h!]
\centering
\includegraphics[scale=1.0, width=8.5cm]{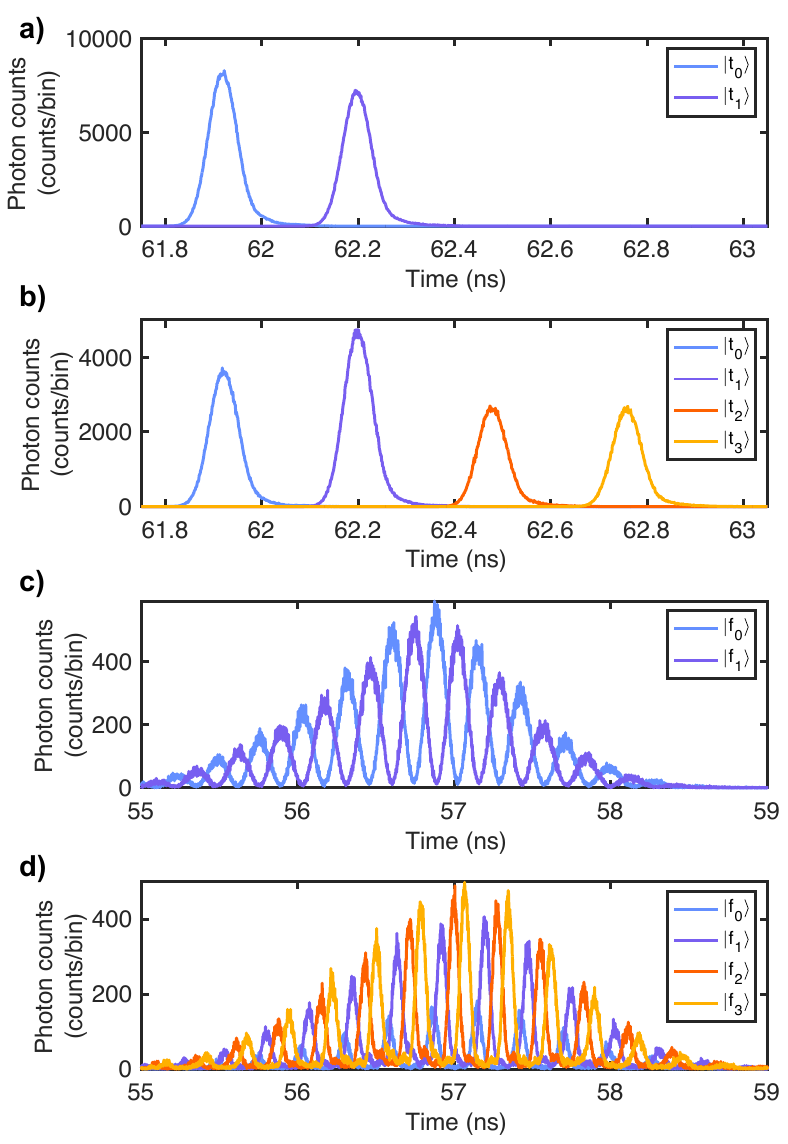}
\caption{
Histograms of two and four-dimensional quantum states used in the QKD experiment measured for the infrastructure node no. 4 at ul.~Ho\.za~69. The states were generated in the pseudo-random order, and the resultant sequence was transmitted for a minute.
a) Z-basis symbols used for the two-dimensional encoding, $Z$-basis measurement; b) $Z$-basis symbols used for the four-dimensional temporal encoding, $Z$-basis measurement; c) $X$-basis symbols used for the two-dimensional encoding, $X$-basis measurement; d) $X$-basis symbols used for the four-dimensional encoding, $X$-basis measurement.
\label{fig:detection}}
\end{figure}

In Fig.~\ref{fig:detection} we show the measured histograms of symbols from the temporal key-generation basis $Z$ used in the experimental realization of the high-dimensional BB84 protocol.
The symbols were $70$~ps wide and separated by $279$~ps, as expected.
The separation between the pulses meets the condition given by Eq.~\eqref{eq:TalbotGDD} for the GDD of the dispersion compensating module and allows distinguishing the $X$-basis symbols using the temporal Talbot effect.
The interference fringes overlap (see Fig. \ref{fig:generation}b), resulting in imperfect distinction and in an increased QBER.
This is the fundamental limit of our method that causes high detection error rate even without the detector timing jitter.
Presence of the detector timing jitter further increases detection error rate \mbox{\cite{widomski2024}}.
It is the dominant component of the $X$ basis QBER.
In Fig.~\ref{fig:detection} we see that the range of detection time in the $X$-basis is approximately $5$~ns for the dimensions two and four.
It is wider than the time occupied by the prepared symbols which may suggest a limitation for clock rate.
But the detection time is shorter than the $10$~ns dead time of the detector, which here is the main clock rate limitation.

\begin{figure}[h!]
\centering
\includegraphics[scale=1.0, width=8.5cm]{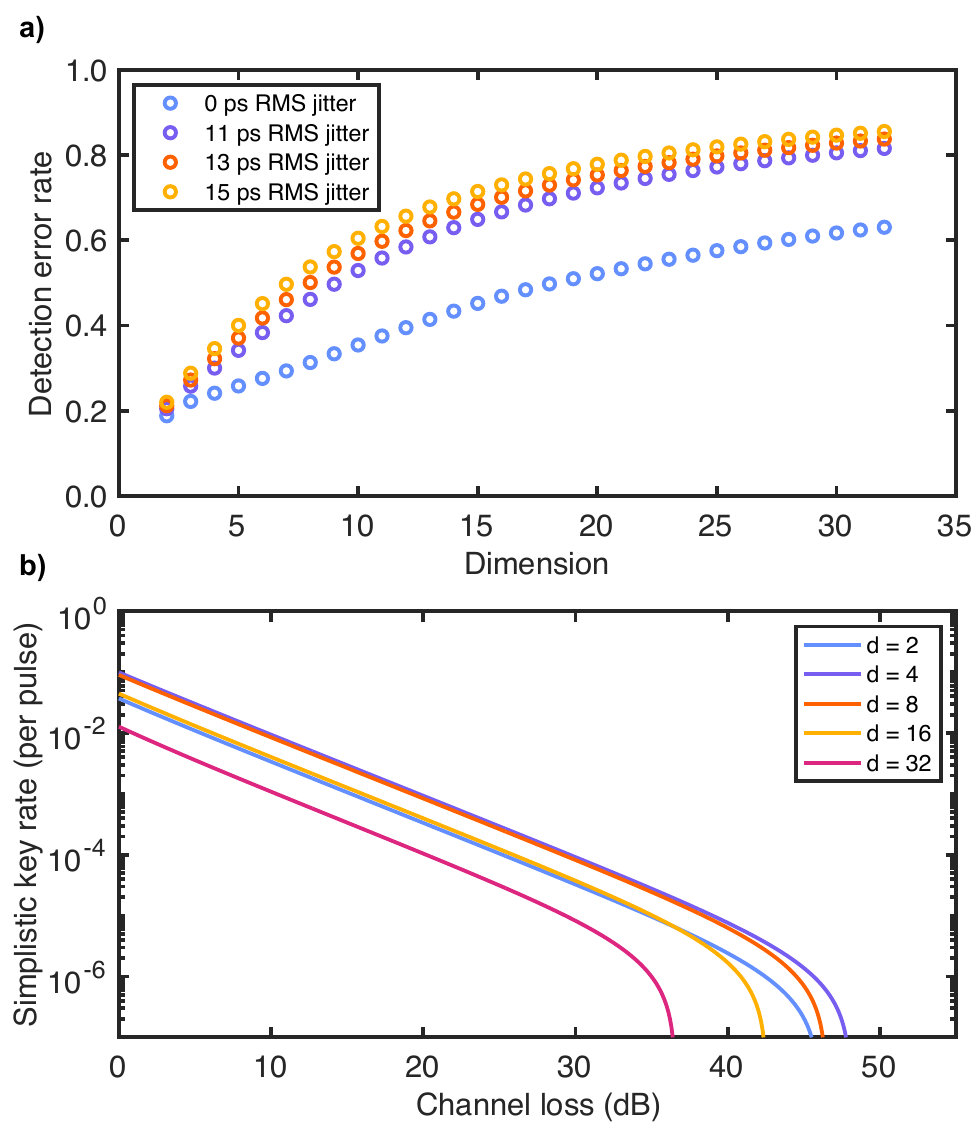}
\caption{
Detection error rate of the temporal Talbot effect based method is the main ingredient of the $X$-basis QBER. It depends on the dimension and the detector timing jitter \cite{widomski2024}. It affects the simplistic key rates obtained with the standard high-dimensional BB84 proof. Hence the optimal dimension is not necessarily the highest dimension allowed by the source and detection timing constraints as usually expected in high-dimensional QKD \cite{hdqkd2019review}.
a) Detection error rates are simulated for different values of jitter and dimensions of encoding.
They are a result of the overlap of probability distributions which is higher for higher dimensions. The overlap in the 2 and 4 dimensional cases is visible in Fig.~\ref{fig:detection}c), \ref{fig:detection}d) and is further discussed in \cite{widomski2024}.
b) Simulated theoretically achievable simplistic key rate values as a function of channel loss for symbols of different dimensions, considering the detection error rate calculated for $15$~ps RMS detection jitter and QBER increase due to the dark count probability. We present relevant simulation parameters in the Supplement.}
\label{fig:qber}
\end{figure}

In Fig.~\ref{fig:qber}a) we present results of a numerical simulation of detection error rate for a range of detection timing jitter values, and for range of dimensions.
Regardless of the relatively high QBER values in the $X$ basis due to the detection errors related to our method, we can still obtain positive key rates using the standard high-dimensional BB84 security proof \cite{sheridan2010}. We use this key rate formula for the analysis further throughout this section to show the relevance of the simple detection method using the temporal Talbot effect and to show superior performance of high-dimensional encoding in this noisy scenario.
However, the detection with a dispersive module may introduce mode-dependent detection efficiency mismatch as explained in section \ref{sec:discussion} and \cite{theorypaper}.
This violates the assumptions of the standard proof and undermines the security of the calculated key rates.
To avoid confusion we will call them the simplistic key rates.

Data collected in the measurement rounds described in the section \ref{setup} was split into calibration data (acquired during the first second of data collection) and experiment data.
For the proof of principle experiment presented here, the calibration data was used to calculate the intensity of the prepared pulse. In an operational QKD system the intensity should be kept constant and be locally verified in the transmitter module.
We included the measured intensities in Section 2 of Supplement 1.

We employed  Eq.~\eqref{eq:skr} to simulate the simplistic key rates, using experimentally obtained parameters.
In Fig.~\ref{fig:qber}b) we present simulated simplistic key rate values for different dimensions and simulated $X$-basis QBERs calculated for $15$~ps RMS detection jitter.
Gains and QBERs are simulated based on channel loss, detection error rates, dark count probabilities as explained in the Appendix E 1 of \cite{theorypaper}.
The values of $X$-basis QBER used for the simulation of the simplistic key rates come mostly from the detection error of the used method and additionally from the influence of the detector's dark counts.
The $Z$-basis detection error rates were experimentally measured for two and four-dimensional symbols in the laboratory and were lower than $0.5\%$. The measured values are provided in Section 1 of Supplement 1.
This value is independent of the dimension due to the nature of time-bin encoding and was assumed for all dimensions for the simulation.
The low $Z$-basis QBER values are stemming from the high extinction ratio of the two cascaded MZMs in our setup and low dark counts of the SNSPDs.
We show the $Z$-basis detection error rate scaling with the extinction ratio and other parameters used for the simulation in Section 1 of Supplement 1.

With a larger alphabet we send more bits of information, but we also witness higher error rates in the control basis as a result of the overlap of the probability density functions.
That trade-off constitutes the limitation, which would further be altered by different values of the detection jitter.
In our scenario four and eight-dimensional states yield the highest key rate, and provide significantly better performances than qubits.

In Fig.~\ref{fig:skr_results} we show experimentally obtained simplistic key rates for qubits and ququarts using Eq.~\eqref{eq:skr} in the case of in-laboratory measurements  and measurements over the dark fiber network.
The measured detection error rate is different from the simulated detection error rate due to the imperfections in the state preparation, detection synchronization, characterization of the detection jitter and total channel dispersion \cite{widomski2024}.
To account for that discrepancy for the simulated simplistic key rates in Fig.~\ref{fig:skr_results} we assume the average detection error rate.
The laboratory measurement results show good consistency with theoretical expectations and low variations of the simplistic secret key rate values between consecutive points.
The infrastructure measurements were conducted over multiple days.
Different conditions caused higher variations of the measured pulse intensity.
These fluctuations may originate from modulator bias imperfections and various stray light sources in the urban fiber network; the measured values are provided in Section~2 of Supplement~1.
The intensities $\mu_i$ were taken as the average over measured intensities over measurement rounds for various attenuations of the channel.
Overall, the pulse intensity for the infrastructure measurements was higher than the pulse intensity for the laboratory measurements.
This resulted in higher simplistic key rates for the infrastructure measurements.
Our experiment served as a proof of principle, and further improvements could be obtained by introducing fast feedback loops for precise correction of the modulator's drift.

Nevertheless, the ququart clearly outperforms the qubit in both the cases in spite of higher $X$-basis QBER, displaying advantages of increased dimensionality.
 
\begin{figure}[h!]
\centering
\includegraphics[scale=1.0, width=8.5cm]{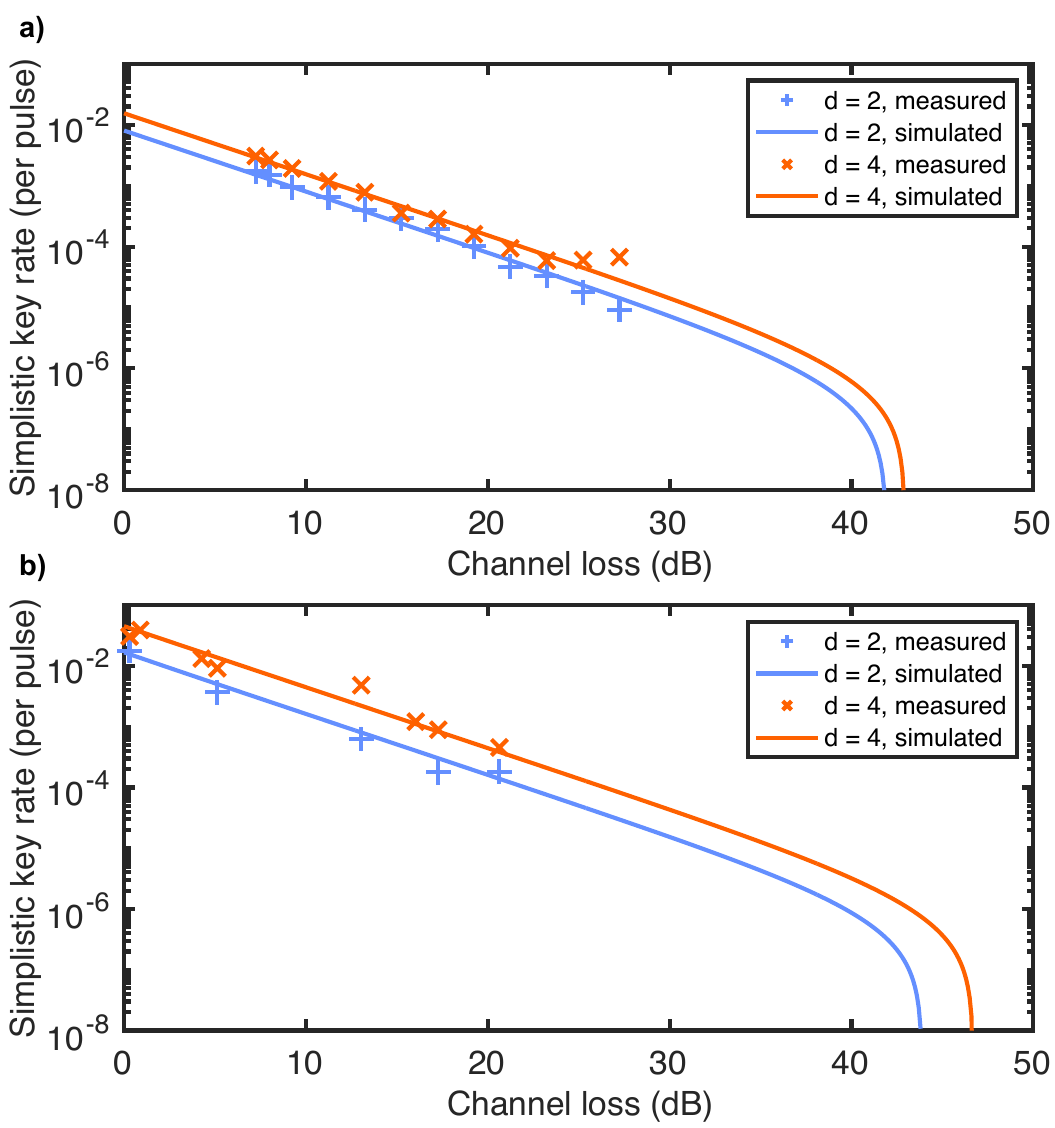}
\caption{Experimentally obtained simplistic key rates per pulse for two and four-dimensional encoding a) In-laboratory measurements including the two optional fiber spools. b) Measurements over dark fiber infrastructure with and without the two fiber spools.
The fluctuations in the pulse intensity cause deviation of the measured simplistic key rates from the simulated values.}
\label{fig:skr_results}
\end{figure}

\section{Security discussion}
\label{sec:discussion}

In the remaining part of the paper we would like to highlight potential security issues pertaining to our setup, and to QKD setups in general, related to basis-dependent detection probabilities. We will discuss our setup in this context and provide an analysis based on new theoretical developments \cite{theorypaper}.

Many QKD protocols allow dropping a measurement round if the receiving side did not register a click in any basis.
The standard decoy-BB84 $d$-dimensional protocol security proof used in this work is one of them, but it assumes that, for any incoming state, the probability of accepting the round is the same for both the measurement bases. That is, the positive operator-valued measure (POVM) elements of the measurements in the $X$ and $Z$ bases corresponding to the rejection of a round are identical.

In our setup, the detectors of the two bases have different efficiencies and, additionally, in the $X$ basis we have an additional insertion loss of the dispersion compensating module.
This violates the assumption of the basis-independent detection efficiency required by standard security proof \cite{theorypaper},
but it could be fixed by inserting an additional attenuator in the $Z$ basis.
In the case of the proof-of-principle data presented in Fig.~\ref{fig:skr_results}, the attenuation was applied in post-processing.
In general, a detection efficiency mismatch may stem from using one or more delay line interferometers or modulators in one of the measurement bases, different efficiencies or qualities of the single-photon detectors, as well as from their varying dark count rate.

However, asymmetric detection efficiencies can also be mode-dependent. For example, single-photon detectors may have different efficiencies depending on the wavelength of the signal.
In some protocols this may be resolved by sufficiently narrow spectral filters placed before the detectors.
Yet, even if the detectors in both the bases in the setup described in section \ref{setup} were identical, the time-frequency mode dependence of the measurements is more subtle.
The dispersion compensating module introduces wavelength-dependent delays, mixing the time and wavelength dependence of the measurement.
This vulnerability is not specific for our setup, but is common for spectro-temporal implementations \cite{Zhong2015, Lee19, Leifgen2015}.

An eavesdropper may perform an intercept-and-resend attack where pulses much shorter than Alice's ones are prepared and sent to Bob.
These pulses will provide correct measurements in the $Z$ basis and discarded rounds in the $X$ basis.
Indeed, in the $X$ basis, group delay dispersion will stretch them much more than Alice's pulses, since the Fourier transform of pulses shorter in time is broader in frequency.
With high probability they will be measured outside the time interval accepted by Bob, leading to no-detection events.
Since discarded measurements do not affect the QBER, the attacker can influence the observed error rate via this mode manipulation and, at the end of the protocol, will share the majority of Bob's bits without being detected, thus compromising the overall security.
Standard QKD proofs are usually only valid given the assumption that the detection probabilities are basis independent, precisely because of attacks such as these.
Therefore, discrepancies between mathematical security proofs and experimental implementations may lead to successful quantum hacking attacks, and are often present in real implementations. For instance, a novel attack on BB84 protocols using asymmetric detectors is discussed in \cite{theorypaper}.

It is not possible to ensure that the detection probability in both bases is perfectly equal and mode-independent in real QKD systems. Indeed,
deviations smaller than  measurement errors in device characterizations could in principle be used by an attacker.
Inclusion of such discrepancies in a security proof of a protocol strengthens the security of the real system. Here we will discuss, in an experimental context, a recently proposed global approach to tackling the basis detection efficiency mismatch issue. 

\section{Tunable beam splitter protocol}
\label{sec:tbs}
\begin{figure}[h!]
\centering
\includegraphics[scale=1.0, width=8.5cm]{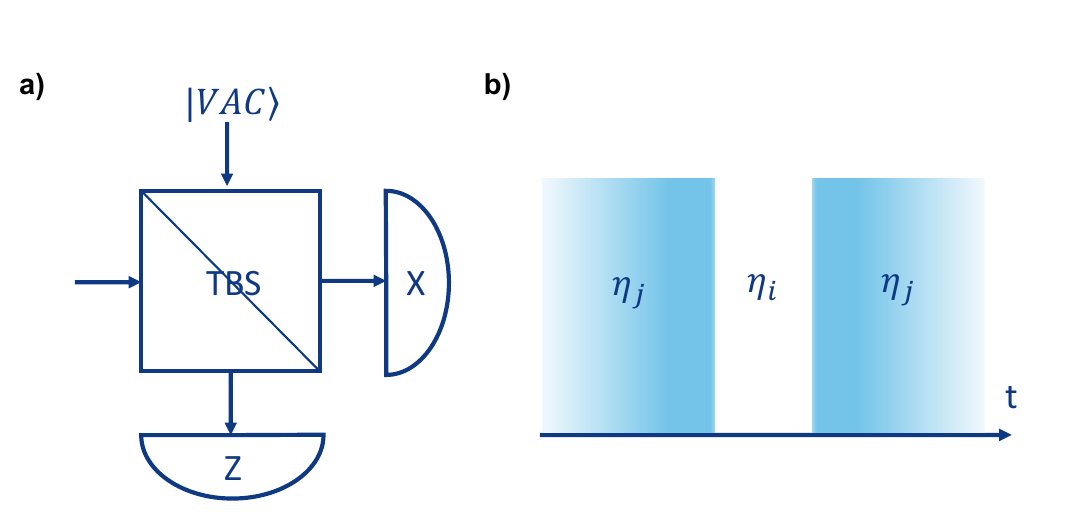}
\caption{Tunable beam splitter (TBS). a) Schematic of a tunable beam splitter used on Bob's side. The incoming signal is split and detected either in the $X$ or the $Z$ basis.
b) Transmission of the TBS is set to a value $\eta_i$ inside the window where the prepared symbol is expected, and to a value $\eta_j$ outside of that interval.
\label{fig:tbs}}
\end{figure}

An analytical security proof dropping the assumption on equal detection efficiency of the two bases is presented in Ref.~\cite{theorypaper}. The protocol assumes using a tunable beam splitter (TBS) on Bob's side -- a device that allows rapidly switching between measurement bases.
The TBS replaces the passive beam splitter in the setup in Fig.~\ref{fig:setup}.
In the protocol this is a realistic device with a finite switching contrast, where maximal and minimal transmission are noted as  $\eta_\downarrow$, $\eta_\uparrow$, and $\eta_2$ is an intermediate state.
The TBS is capable of switching its transmission from value $\eta_j$ to $\eta_i$ and back to $\eta_j$, where the time interval in which the transmission is equal to $\eta_i$ coincides with the $Z$ basis time window.
Detection settings $\eta_j, \eta_i$ are randomly chosen (cf. Fig.~\ref{fig:tbs}) to obtain detection gains corresponding to those settings.
Then the detector decoy technique \mbox{\cite{detector-decoy}} is used to link measured detection statistics to estimate the probability that a state would be detected in the basis $Z$ but rejected in the basis $X$ \cite{theorypaper}.
For the security proof to follow, some assumptions on the three possible transmissions need to be satisfied:

\begin{align}
    \eta_\uparrow &> \frac{\eta_\downarrow}{1-\eta_\downarrow} \label{constraint1}, \\
    \frac{\eta_X}{\eta_Z} &> (\eta_\downarrow)^{-1} \left(1-\sqrt{1-\frac{\eta_\downarrow}{\eta_\uparrow}}\right) \label{constraint2},\\
    \eta_\downarrow&<\eta_2<\eta_\uparrow \label{constraint3}.
\end{align}

In our analysis we use a close-to-optimal choice of the intermediate transmission: $\eta_2 = (1/4)(\sqrt{\eta_\downarrow} + \sqrt{\eta_\uparrow})^2$. The TBS could be implemented for example with a double output Mach-Zehnder modulator (MZM) \cite{Leifgen2015}. This kind of MZM is also available within generic photonic integrated technology, and was used to match detection efficiencies \cite{sibson2017chip}.

\begin{figure}[h!]
\centering
\includegraphics[scale=1.0, width=8.5cm]{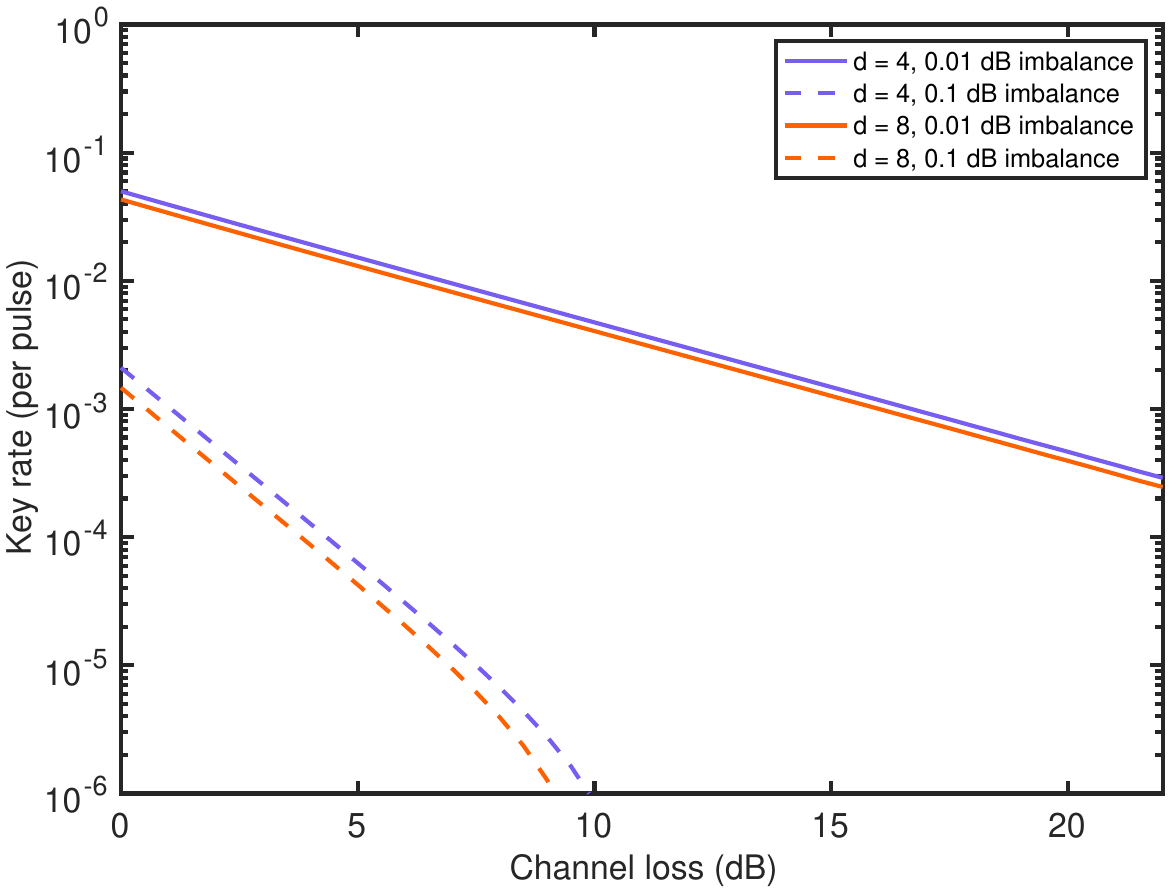}
\caption{Simulated key rate per pulse values considering 0.1 dB and 0.01 dB efficiency mismatch obtained according to the security proof provided in \cite{theorypaper}. Key rate values were optimized over intensity $\mu_1$ with fixed $\mu_2=2\times10^{-6}$, $\mu_3=1\times10^{-6}$ (optimal values of $\mu_1$ are provided in Section 2 of Supplement 1).
\label{fig:skr_mismatch}}
\end{figure}

The resultant key rate can be obtained with the aforementioned Eq.~\eqref{eq:skr} by applying expressions for phase error rate, yields, QBERs, and gains described in \cite{theorypaper}.
One of the main outcomes of the security proof of \cite{theorypaper} is the derivation of the phase error rate upper bound, which is obtained without the usual assumption on equal detection efficiencies in the two bases.
In Fig.\ \ref{fig:skr_mismatch} we present simulated key rate values for $0.1$~dB and $0.01$~dB measurement basis efficiency imbalance.
The resultant key rate values are very sensitive to the efficiency imbalance and to the jitter due to the properties of our detection method.
For our set of experimental parameters the resulting key rate would be negative, which shows how dangerous the efficiency mismatch can be.
Negative values mostly stem from high attenuation difference caused by the DCM and technical limitations of our sender (Alice).
The problem of imbalance can be effectively mitigated by adding an attenuator to one of the measurement bases in order to equalize the efficiencies.
Attenuators with $0.1$~dB accuracy are easily available and precise matching can be achieved with an EVOA, which makes this a convenient and cost-efficient fix.
This shows that the more rigorous proof \cite{theorypaper} will produce positive key rates with the appropriate modifications of the setup.
This shows that the more rigorous proof \cite{theorypaper} will produce positive key rates with the appropriate modifications of the setup.

\section{Conclusion}

In summary, we demonstrate a high-dimensional BB84 QKD protocol with two and four-dimensional symbols in an all-fiber setup using one single-photon detector per measurement basis. We successfully employ the temporal Talbot effect to detect high-dimensional phase-encoded superpositions by mapping frequency to time. 
This method uses only one detector to detect states from the control basis by the measurement of the time of arrival of the photon.
We obtain positive simplistic key rates using the standard HD BB84 security proof in spite of the relatively high QBER, which is stemming from the overlap of the probability distributions and is strongly dependent on the detection timing jitter.
We provide a comparison of theoretically-achievable simplistic key rates for a variety of dimensions based on a simulation with experimentally-obtained parameters. We show how basis efficiency mismatch and the security proof can severely impact the resultant key rate. We show vastly different key rates where differences stemmed from the considerations of the potential eavesdropper attacks within the security-proof framework.
Our method is limited by low (sub-gigahertz) repetition rates, and the fact that the resultant shape of the superposition in the $X$ basis occupies numerous time bins. 
Our measurement setup is prone to mode-induced detection efficiency mismatches, which can be accounted for by the security proof developed in \cite{theorypaper}, in combination with introduction of a tunable beam splitter and optical attenuators at the receiver side. In this regard, optical attenuators are easily obtainable, and tunable beam splitters can be realized as double output MZMs.

Given the properties of the temporal Talbot effect, either the symbol separation can be digitally adjusted with respect to the dispersion of the quantum channel and receiver, or the GDD value at the receiver side can be selected to meet the requirements.
This technique supports pulses of full width at half maximum (FWHM) of up to $100$~ps, which makes using generically-available photonic integrated circuits (PICs) feasible for constructing transmitter modules \cite{widomski_pic}.
In general, the  visibility of the temporal Talbot effect depends on the relation of width of a single pulse to the time separation of pulses and the number of pulses in the superposition \cite{Azana:finite:pulse}.
Optimization over these parameters for the QKD protocols is an interesting direction of further research.
Photonic integration of the receiver would require using on-chip GDD media and detectors, which is currently an emerging solution \cite{yu2022, trenti2022}. In spite of that, PICs may enable integration and deployments due to available bandwidth, small form factor, cost-effectiveness, and the possibility of constructing double output MZMs \cite{sibson2017chip, Paraiso2021}.
Our proposal offers low complexity and high flexibility, which is crucial for quantum-photonic applications. Our findings may be an interesting consideration for constructing high-dimensional systems as well as deployment and development of qubit-based solutions, since the aforementioned basis efficiency mismatch problem is independent of the encoding dimension.
\begin{backmatter}

\bmsection{Funding}
Federico Grasselli contributed to this work exclusively on behalf of Heinrich-Heine-Universit\"at D\"usseldorf (previous affiliation).
A part of this work was supported by the  QuICHE Project, which is supported by the National Science Center, Poland (project no.~2019/32/Z/ST2/00018) under QuantERA, which has received funding from the European Union's Horizon 2020 research and innovation programme under grant agreement no.~731473.
Maciej Ogrodnik, Adam Widomski and Micha{\l} Karpi\'nski appreciate funding from the University of Warsaw within the ``Excellence Initiative –- Research University'' framework.
Hermann Kampermann, Dagmar Bru\ss{} and Nikolai Wyderka acknowledge support by the QuantERA project QuICHE via the German Ministry of Education and Research (BMBF grant no.~16KIS1119K).
Dagmar Bru\ss{} and Hermann Kampermann acknowledge support by the German Ministry of Education and Research through the project QuNET+ProQuake (BMBF grant no.~16KISQ137) and QuKuK (BMBF grant no.~16KIS1619).
Giovanni Chesi and Chiara Macchiavello acknowledge the EU H2020 QuantERA ERANET Cofund in Quantum Technologies project QuICHE and support from the PNRR MUR Project PE0000023-NQSTI.
Nathan Walk acknowledges funding from the BMBF (QPIC-1, Pho-Quant, QR.X).

\bmsection{Acknowledgments}
We thank P.~Rydlichowski and P.~Celmer for assistance during the measurements using the fiber-optic infrastructure.

\bmsection{Disclosures}
The authors declare no conflicts of interest.

\bmsection{Data availability}
Data underlying the results presented in this paper can be obtained from the authors upon a reasonable request.

\bmsection{Supplemental document}
See Supplement 1 for supporting content.

\end{backmatter}

\bibliography{sample}

\onecolumn

\appendix
\renewcommand{\thesection}{\arabic{section}}
\section*{Supplement 1}
\section{Errors in detection}

\begin{figure}[htbp]
\centering
\includegraphics[scale=1.0, width=8.5cm]{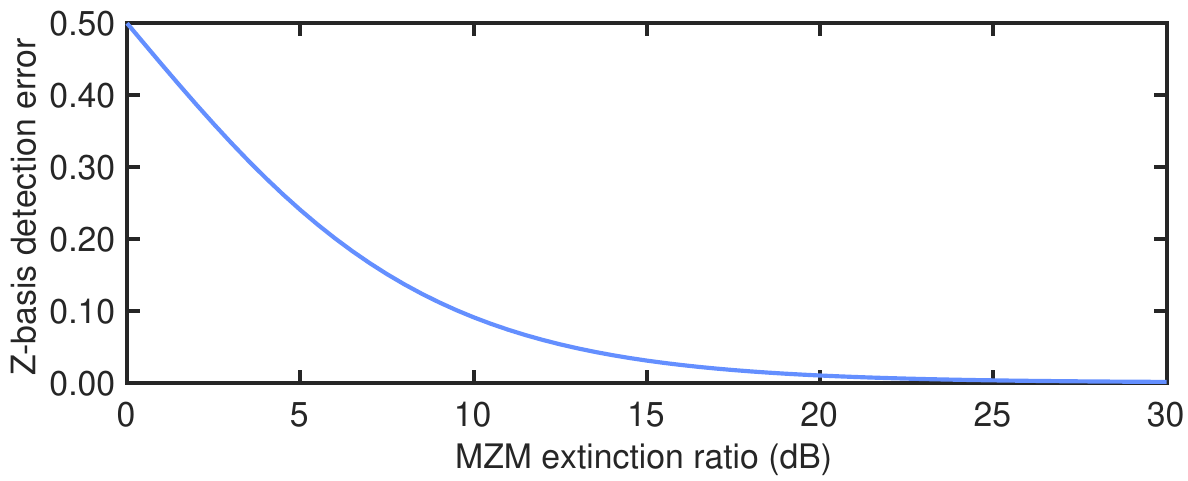}
\caption{$Z$-basis detection error scaling with modulator's extinction ratio.
\label{fig:zqber}}
\end{figure}

The $Z$-basis detection error rate is dependent on the extinction ratio of the Mach-Zehnder modulator (MZM). It is given by \footnote{D. Scalcon, E. Bazzani, G. Vallone, et al., “Low-error encoder for time-bin and decoy states for quantum key
distribution,” npj Quantum Inf. 11, 22 (2025).}:
\begin{equation}
    ERROR_Z=\frac{1}{1+10^\frac{ER}{10}},
\end{equation}
where extinction ratio $ER$ is expressed in dB.
The detection error values corresponding to a range of extinction ratios spanning up to $30$ dB are presented in Fig.~\ref{fig:zqber}.

Extinction ratios higher than $20$ dB result in detection error rate values lower than $1\%$.
The typical extinction ratio of MZMs is $40$~dB.
Even if full dynamic range is not used during modulation, the resultant pulse contrast is high enough for QKD application, assuming that the modulator was biased for extinction to minimize light leak.
We used two cascaded MZMs biased for extinction in our setup to generate pulses and to achieve high light leak suppression.
For simulations in Fig.~5b) and Fig.~8 we use detection error rate $ERROR_Z=0.5\%$, which is higher than the detection error rates measured during laboratory experiments (see table \ref{table:S1}).

\begin{table}[H]
\centering
\caption{\bf In-laboratory measurements of QBER for two and four-dimensional symbols} \label{table:S1}
\begin{tabular}{c|cc|cc}
\hline
    \multirow{2}{*}{Attenuation (dB)} &
      \multicolumn{2}{c|}{Dimension 2} &
      \multicolumn{2}{c}{Dimension 4} \\ \cline{2-5}
      & $QBER_X\ (\%)$ & $QBER_Z\ (\%)$ & $QBER_X\ (\%)$ & $QBER_Z\ (\%)$\\
\hline
$7.24 $  & $21.83$ & $0.208$ & $37.53$ & $0.309$\\
$8.00 $  & $21.80$ & $0.188$ & $35.94$ & $0.317$\\
$9.24 $  & $21.96$ & $0.205$ & $36.82$ & $0.331$\\
$11.24$  & $21.77$ & $0.202$ & $35.83$ & $0.303$\\
$13.24$  & $21.64$ & $0.201$ & $35.98$ & $0.319$\\
$15.24$  & $21.69$ & $0.210$ & $35.65$ & $0.316$\\
$17.24$  & $21.63$ & $0.203$ & $35.40$ & $0.369$\\
$19.24$  & $21.76$ & $0.289$ & $35.43$ & $0.443$\\
$21.24$  & $21.88$ & $0.295$ & $34.95$ & $0.498$\\
$23.24$  & $22.04$ & $0.469$ & $42.67$ & $0.587$\\
$25.24$  & $21.83$ & $0.544$ & $36.80$ & $1.000$\\
$27.24$  & $23.67$ & $0.599$ & $36.69$ & $0.991$\\
\hline
\end{tabular}
\end{table}

To plot the simulated curves in Fig. 6a) (in-laboratory data) we assumed average values from the table \ref{table:S1}: $\text{QBER}_Z=0.301\%$ and $\text{QBER}_X=21.96\%$ for $d=2$, and $\text{QBER}_Z=0.482\%$ and $\text{QBER}_X=36.65\%$ for $d=4$ as the detection error rates.
To plot the simulated curves in Fig. 6b) (infrastructure data) we assumed average values from the table \ref{table:S2}: $\text{QBER}_Z=0.529\%$ and $\text{QBER}_X=24.50\%$ for $d=2$, and $\text{QBER}_Z=0.541\%$ and $\text{QBER}_X=35.64\%$ for $d=4$ as the detection error rates.
Detection error rate values in the $X$-basis used for both Fig.~5b) and Fig.~8 are displayed in Fig.~5a).
In particular the simulated values of the $X$-basis detection error rate were: $ERROR_X=21.97\%$ for $d=2$, and $ERROR_X=34.56\%$ for $d=2$.
The measured values of $QBER_X$ differ from the simulated detection error rate due to the imperfections in the state preparation, detection synchronization, characterization of the detection jitter and total channel dispersion that are not fully modeled in the simulation.

\begin{table}[H]
\centering
\caption{\bf Infrastructure measurements of QBER for two and four-dimensional symbols} \label{table:S2}
\begin{tabular}{c|cc|cc}
\hline
    \multirow{2}{*}{Attenuation (dB)} &
      \multicolumn{2}{c|}{Dimension 2} &
      \multicolumn{2}{c}{Dimension 4} \\ \cline{2-5}
      & $QBER_X\ (\%)$ & $QBER_Z\ (\%)$ & $QBER_X\ (\%)$ & $QBER_Z\ (\%)$\\
\hline
$0.30 $ & $22.60$ & $0.0611$ & $39.11$ & $0.0977$\\
$0.89 $ & $23.28$ & $0.0553$ & $37.18$ & $0.101 $\\
$4.27 $ & -       & -        & $36.30$ & $0.270 $\\
$5.13 $ & $23.26$ & $0.297 $ & $35.46$ & $0.315 $\\
$13.04$ & $24.00$ & $1.19  $ & $35.78$ & $0.240 $\\
$16.04$ & $27.54$ & $0.431 $ & $34.03$ & $0.501 $\\
$17.28$ & $26.76$ & $0.676 $ & $33.41$ & $1.04  $\\
$20.64$ & $24.08$ & $0.988 $ & $33.88$ & $1.77  $\\
\hline
\end{tabular}
\end{table}

\section{Pulse intensities}
\label{supp:sec:intensities}

\begin{table}[H]
\centering
\caption{\bf In-laboratory measurements of the mean photon number for two and four-dimensional symbols} \label{table:S3}
\begin{tabular}{c|ccc|ccc}
\hline
    \multirow{2}{*}{Attenuation (dB)} &
      \multicolumn{3}{c|}{Dimension 2} &
      \multicolumn{3}{c}{Dimension 4} \\ \cline{2-7}
      & $\mu_1$  & $\mu_2$ & $\mu_3$ & $\mu_1$  & $\mu_2$ & $\mu_3$\\
\hline
$7.24 $ & $0.0635$ & $0.503\times10^{-3}$  & $48.1\times10^{-6}$ & $0.0602$ & $0.510\times10^{-3}$  & $42.8\times10^{-6}$\\
$8.00 $ & $0.0672$ & $0.639\times10^{-3}$  & $42.4\times10^{-6}$ & $0.0624$ & $0.487\times10^{-3}$  & $30.6\times10^{-6}$\\
$9.24 $ & $0.0610$ & $0.515\times10^{-3}$  & $46.1\times10^{-6}$ & $0.0603$ & $0.546\times10^{-3}$  & $42.6\times10^{-6}$\\
$11.24$ & $0.0610$ & $0.454\times10^{-3}$  & $35.2\times10^{-6}$ & $0.0635$ & $0.445\times10^{-3}$  & $76.7\times10^{-6}$\\
$13.24$ & $0.0658$ & $0.604\times10^{-3}$  & $73.5\times10^{-6}$ & $0.0619$ & $0.508\times10^{-3}$  & $60.0\times10^{-6}$\\
$15.24$ & $0.0669$ & $0.612\times10^{-3}$  & $74.6\times10^{-6}$ & $0.0626$ & $0.478\times10^{-3}$  & $86.8\times10^{-6}$\\
$17.24$ & $0.0663$ & $0.562\times10^{-3}$  & $117 \times10^{-6}$ & $0.0619$ & $0.530\times10^{-3}$  & $123 \times10^{-6}$\\
$19.24$ & $0.0605$ & $0.526\times10^{-3}$  & $126 \times10^{-6}$ & $0.0600$ & $0.607\times10^{-3}$  & $155 \times10^{-6}$\\
$21.24$ & $0.0655$ & $0.791\times10^{-3}$  & $200 \times10^{-6}$ & $0.0632$ & $0.587\times10^{-3}$  & $224 \times10^{-6}$\\
$23.24$ & $0.0631$ & $0.783\times10^{-3}$  & $331 \times10^{-6}$ & $0.0627$ & $0.845\times10^{-3}$  & $327 \times10^{-6}$\\
$25.24$ & $0.0642$ & $1.130\times10^{-3}$  & $457 \times10^{-6}$ & $0.0604$ & $0.989\times10^{-3}$  & $480 \times10^{-6}$\\
$27.24$ & $0.0625$ & $1.090\times10^{-3}$  & $567 \times10^{-6}$ & $0.0611$ & $1.070\times10^{-3}$  & $784 \times10^{-6}$\\
\hline
\end{tabular}
\end{table}

For theoretical simulation results in Fig. 5b) (BB84 protocol) the $\mu_1$ value was found during an optimization process.
Low values of $\mu_2$ and $\mu_3$ are optimal for the decoy state method \cite{theorypaper}.
Chosen values reflect the experimentally achievable intensities due to the finite extinction ratio of MZMs.
The achievable key rate values are always lowered by the detector's dark counts. To account for this factor, we measure the probability that a dark count will contribute to the outcome in the $X$ ($p_{dcX}$) and $Z$ bases ($p_{dcZ}$). Those probabilities differ due to internal properties of the individual detector's channels. The simulation parameters were: $d=2,4,8,16,32$ and the corresponding $\mu_1=0.65, 0.77, 0.76, 0.62, 0.39$ respectively.
For all dimensions we use $\mu_2=2\times10^{-6}$, $\mu_3=1\times10^{-6}$, $\eta_x=84\%$, $\eta_z=81\%$, $p_{dcX}=3.36\times10^{-7}$, $p_{dcZ}=2.80\times10^{-7}$.

For Fig. 6 we use the average of intensities measured in the calibration rounds of the protocol for various attenuations.
For Fig.~6a) (in-laboratory data) we use $\mu_1=0.064$, $\mu_2=0.684\times10^{-3}$, $\mu_3=177\times10^{-6}$ for $d=2$, and $\mu_1=0.0617$, $\mu_2=0.634\times10^{-3}$, $\mu_3=203\times10^{-6}$ for $d=4$ to plot the theoretical curves and compute experimental key rates for a range of attenuations. The exact $\mu$ values for each measurement point are presented in table \ref{table:S3}.

For Fig. 6b) (infrastructure data) we use $\mu_1=0.236$, $\mu_2=2.6\times10^{-3}$, $\mu_3=355\times10^{-6}$ for $d=2$, and  $\mu_1=0.2$, $\mu_2=2.49\times10^{-3}$, $\mu_3=329\times10^{-6}$ for $d=4$ to plot the theoretical curves and compute experimental key rates for a range of attenuations. The exact $\mu$ values for each measurement point are presented in table \ref{table:S4}.

\begin{table}[H]
\centering
\caption{\bf Infrastructure measurements of the mean photon number for two and four-dimensional symbols} \label{table:S4}
\begin{tabular}{c|ccc|ccc}
\hline
    \multirow{2}{*}{Attenuation (dB)} &
      \multicolumn{3}{c|}{Dimension 2} &
      \multicolumn{3}{c}{Dimension 4} \\ \cline{2-7}
      & $\mu_1$  & $\mu_2$ & $\mu_3$ & $\mu_1$  & $\mu_2$ & $\mu_3$\\
\hline
$0.30 $ & $0.105$ & $1.57\times10^{-3}$ & $64.6\times10^{-6}$ & $0.103$ & $1.4 \times10^{-3}$ & $92.3\times10^{-6}$ \\
$0.89 $ & $0.153$ & $0.78\times10^{-3}$ & $96.8\times10^{-6}$ & $0.152$ & $2.17\times10^{-3}$ & $123 \times10^{-6}$ \\
$4.27 $ & -       & -                  &  -                 & $0.107$ & $2.73\times10^{-3}$ & $67.3\times10^{-6}$ \\
$5.13 $ & $0.102$ & $0.86\times10^{-3}$ & $32.1\times10^{-6}$ & $0.104$ & $0.86\times10^{-3}$ & $40.9\times10^{-6}$ \\
$13.04$ & $0.462$ & $6.98\times10^{-3}$ & $522 \times10^{-6}$ & $0.426$ & $4.03\times10^{-3}$ & $481 \times10^{-6}$ \\
$16.04$ & $0.271$ & $1.31\times10^{-3}$ & $169 \times10^{-6}$ & $0.226$ & $1.95\times10^{-3}$ & $195 \times10^{-6}$ \\
$17.28$ & $0.252$ & $2.47\times10^{-3}$ & $592 \times10^{-6}$ & $0.233$ & $2.69\times10^{-3}$ & $549 \times10^{-6}$ \\
$20.64$ & $0.307$ & $4.20\times10^{-3}$ & $1010\times10^{-6}$ & $0.253$ & $4.09\times10^{-3}$ & $1090\times10^{-6}$ \\
\hline
\end{tabular}
\end{table}

The simulation parameters for the theoretical simulation results in Fig. 8 (TBS protocol) were: $\mu_2=2\times10^{-6}$, $\mu_3=1\times10^{-6}$, $\eta_x=\eta_z=84\%$, $\eta_\downarrow=0.0001$, $\eta_2=0.2549$, $\eta_\uparrow=0.9999$, $p_{d\,cX}=p_{d\,cZ}=2.8\times10^{-7}$. Selected $\eta_\uparrow$ and $\eta_\downarrow$ values correspond to $40$~dB ER of a TBS. Attenuation in the $X$ basis was given by the insertion loss of the DCM, and was equal to $2.67$~dB. We numerically set the attenuation in the $Z$ basis to $2.66$~dB ($2.57$~dB) to show the effect of measurement efficiency mismatch for $d=4$ ($8$). We selected those dimensionalities as the resultant key rate values are the highest considering our detection method. For $d=4$ we use $\mu_1=28.58\times10^{-3}$ and $\mu_1=0.79$ for $0.1$ dB and $0.01$ dB imbalance respectively. For $d=8$ we use $\mu_1=20.60\times10^{-3}$ and $\mu_1=0.75$ for the same imbalances.

\end{document}